\PassOptionsToPackage{pdfpagelabels=false}{hyperref} 
\expandafter\def\csname ver@fixltx2e.sty\endcsname{}
\documentclass[useAMS,usenatbib]{mnras}

\usepackage{graphicx,epsfig,subfigure}
\usepackage{graphicx}
\usepackage{url}
\usepackage{amsmath}
\usepackage{amssymb}
\usepackage{amsfonts}
\usepackage{bm}
\usepackage{ulem}

\usepackage{float}

\usepackage{rotating}

\usepackage{longtable}

\title[Virial masses of LTGs]
   {Virial masses of late type galaxies from the SDSS DR16}

\author[Nigoche-Netro et al.]
{
A. Nigoche-Netro$^1$\thanks{E-mail: alberto.nigoche@academicos.udg.mx
eduardo.delafuente@academicos.udg.mx}, E. de la Fuente$^{2,3}$$^{\textcolor{blue}{*}}$, R. J. Diaz$^{4,5}$, M. P. Ag\"uero$^{5,6}$,
\newauthor  S. N. Kemp$^1$, R. A. Marquez-Lugo$^1$, P. Lagos$^7$, A. Ruelas-Mayorga$^8$,
\newauthor N. L. López-Contreras$^1$\\
    $^1$Instituto de Astronom\'ia y Meteorolog\'ia, CUCEI, Universidad de Guadalajara, Guadalajara, Jal. 44130, M\'exico. \\
    $^2$Departamento de F\'isica, CUCEI, Universidad de Guadalajara, Guadalajara, Jal. 44130, M\'exico. \\
    $^3$Institute for Cosmic Ray Research, University of Tokyo, Kashiwanoha campus, Chiba, Japan (Sabbatical 2021). \\
    $^4$Gemini Observatory, NSF's NOIRLab, 950 N Cherry Ave, Tucson AZ, USA. \\
    $^5$Universidad Nacional de C\'ordoba, Laprida 854, C\'ordoba, CPA: X5000BGR, Argentina. \\
    $^6$Consejo Nacional de Investigaciones Cient\'ificas y T\'ecnicas (CONICET), Argentina. \\
    $^7$Instituto de Astrof\'isica e Ci\^encias do Espa\c{c}o, Universidade do Porto, CAUP, Rua das Estrelas, 4150-762 Porto, Portugal. \\
    $^8$Instituto de Astronom\'ia, Universidad Nacional Aut\'onoma de M\'exico, Cd. Universitaria, M\'exico, D.F. 04510, M\'exico. 
}

\date{Accepted 2018. Received 2018; in original form 2018}

\pubyear{2018}

\DeclareUnicodeCharacter{2212}{-}

\begin{document}
\label{firstpage}
\pagerange{\pageref{firstpage}--\pageref{lastpage}}
\maketitle

\begin{abstract}
Motivated by the challenges of calculating the dynamical masses of late-type galaxies (LTGs) and the enormous amount of data from the Sloan Digital Sky Survey (SDSS), we calculate virial masses of a sample of approximately 126,000 LTGs from the sixteenth data release of the SDSS. The virial mass estimations were made considering Newtonian mechanics, virial equilibrium and velocity dispersion from stars and gas. The procedure gave as a result seven mass estimations for each galaxy. The calculated masses were calibrated using a sample of spiral galaxies with velocity rotation curves. Considering the results from the calibration, we find that the correlation between virial and dynamical (rotation curve) masses is stronger for high inclination values. Therefore the calibration relies more on the available data for higher inclination angle galaxies. We also show that if we have a heterogeneous sample of galaxies one must take into consideration the size and colour of these galaxies by using the following variables: Sersic index $n$, concentration index and colour of the stars. For relatively smaller and bluer LTGs the gas velocity dispersion provides a more consistent mass calculation, while for LTGs that are relatively larger and redder the stellar velocity dispersion provides a better correlated mass calculation.

\end{abstract}



\section{Introduction}
\label{intro}

The determination of the masses of galaxies has been an interesting problem ever since galaxies were identified as very large conglomerations of stars and gas in the universe \citep[see][]{Burbidge1975,Fich1991,Courteauetal2014}. 
At present there are three main methods that allow the calculation of galactic masses, namely:

\begin{enumerate}

\item  Dynamical or virial masses. The dynamical or virial masses are determined recurring to dynamical methods which in turn use the rotation curves or the velocity dispersion of gas or stars in galaxies. It is important to mention at this point that all these methods presuppose that Newtonian gravitation is valid at all scales in the universe; if this were not the case, then the determinations obtained from these methods would be flawed. 
The dynamical mass determinations for LTGs are usually derived from their rotation curves. Early attempts to determine the masses of galaxies using this method were undertaken by \citet{Scheiner1899}, \citet{Slipher1914}, \citet{Pease1918}, and \citet{Opik1922}, in which he inferred a value for the mass of M31. Some time later there were attempts at determining the mass of the Milky Way by \citet{KapteynvanRhijn1922}, \citet{Oort1932a}, and \citet{Oort1932b}. Determinations of the M31 total mass using velocities determined from absorption lines were made by \citet{Babcock1939}, \citet {Mayall1950}, and \citet{Lallemand1960}. With the measurement of rotation curves further away from the centre of LTGs, and the discovery that these remain flat out to large radii, early suggestions of the existence of a non-luminous massive component were put forward. \citet{RubinFord1970} were the first to obtain a rotation curve for M31 out to $120'$ $(\sim 27 \  kpc)$ in which they noticed how it remains flat out to a large distance from the centre. Using $HI$, \citet{RobertsWhitehurst1975} confirmed the flatness of the M31 rotation curve out to  $170'$ $(\sim 38 \ kpc)$. The flatness of rotation curves in all galaxy types is a well-established fact \citep[][among others]{FaberGallagher1979,Rubinetal1985,SofueRubin2001}. There have also been mass determination of LTGs using the emission lines of $H\alpha$, $CO$ and $HI$. A reasonable agreement between determinations performed with these different lines is found \citep[see][]{SofueRubin2001,Simonetal2003,Simonetal2005,SpekkensSellwood2007}), and also from measurements made with the $[{OII]}$, $[OIII]$, $H \beta$ and $[SII]$ lines \citep{CourteauSohn2003}. At some point the question of whether the gas rotation curve really represented the total mass of the galaxy arose. There is likely evidence that the gas rotation curves represent the gas distribution within the optical disk of galaxies \citep[see][among others]{Cayatteetal1994,MathewsonFord1992,Courteau1997,CatinellaHaynesG2007}.

To determine the mass of an elliptical galaxy using the virial theorem, three things are needed:

\begin{itemize}
    \item Distance of the galaxy from the Sun
    \item the line-of-sight velocity dispersion of the stars in the centre
    \item the distribution of light projected on the plane of the sky from which we can derive the potential energy
\end{itemize}

The first determination of a velocity dispersion was for $M32$ by \citet{Minkowsky1954}, later \citet{BurbidgeBurbidgeF1961a} and \citet{BurbidgeBurbidgeF1961b} reevaluated the value of the velocity dispersion and in \citet{BurbidgeBurbidgeF1961c} they obtained the velocity dispersion for $NGC \ 3379$. Velocity dispersions for 12 additional galaxies were obtained by \citet{Minkowsky1961}.
For the determination of the potential energy of a spheroidal galaxy we refer to \citet{Poveda1958} where a full discussion is given. It was shown by \citet{deVaucouleurs1963} and \citet{Poveda1958} that the potential energy is given by:
\begin{equation}
    \Omega=-0.33\frac{GM^2}{R'}
    \label{eq:Poveda}
\end{equation}

where $R'$ is the radius that contains half of the total light of the galaxy, this radius is called the `effective' radius, $G$ is the constant of Universal Gravitation and $M$ represents the mass. All the quantities in equation (\ref{eq:Poveda}) are in $cgs$ units.

The dynamical masses of Gas-Poor Galaxies $(ETGs)$ have been determined from molecular gas observations by \citet {SageWelchYoung2007} and \citet{Youngetal2011}, from ionised gas observations by \citet{Bertolaetal1984}, \citet{Fisher1997} and \citet{Sarzietal2006} and from neutral gas observations by \citet{KnappTurnerC1985}, \citet{Morgantietal2006}, and \citet{diSeregoetal2006} among others. \citet{KingMinkowsky1966} measured rotation in the inner regions of two giant ellipticals in the Virgo cluster, NGC 4621 (E3) and NGC 4697 (E5).

\item  Luminous masses. Luminous masses of galaxies are determined measuring the total luminosity of galaxies and assuming a specific mass to light $(M/L)$ ratio which would imply a value for the mass. The $(M/L)$  does not have a universal value and its determination is central to the calculation of galaxy masses using this method. The $(M/L)$ ratio may depend on a variety of factors such as the way the galactic mass is assembled \citep{deLuciaetal2007}, the different Spectral Energy Distributions (SED) of galaxies \citep{Walcheretal2011} and \citep{Conroy2013}, the different mix of Stellar Populations (SPs). \citet{Oort1926} and \citet{Baade1944} recognized the presence of at least two different SPs in our galaxy. For a comprehensive review on SPs see \citet{GreggioRenzini2011}. The different SPs are superposed in a galaxy, and each has a different Star Formation Rate (SFR) which must also be considered and which evolves with time  \citep{Sandage1986a}, \citep{MacArthuretal2004}. Another factor is the Initial Mass Function 
\citep[IMF; see][]{Salpeter1955,Scalo1986,Kroupa2001,Chabrier2003} which tells us how the mass is distributed in a galaxy at the beginning of the lives of its stars.

The types of stars in a galaxy -hence their mass- can be calculated using stellar population synthesis models. There are optimised population synthesis models \citep[see][among others] {SpinradTaylor1971,Faber1972,Oconnell1976,Pickles1985,BicaAlloin1986,MacArthurGonzalez2009} and Evolutionary population synthesis models \citep[see][among others]{Tinsley1972,TinsleyGunn1976,RenziniVoli1981,Bruzual1983}. In summary, determining the $M/L$ ratio for a galaxy could be a very complicated problem, however once it is determined, the galactic mass can be calculated straightforwardly from its measured luminosity and distance.

\item  Mass determination using relativistic light bending by gravitational lenses. This method would give us a crude value for the mass of a single galaxy because in order to bend light gravitationally it is necessary to have a very large mass such as that of a cluster of galaxies and the mass of an individual galaxy would correspond to the average value of the mass for each galaxy in the cluster.

The first attempt to detect the weak gravitational bending of light by a massive object was conducted by \citet{Tysonetal1984} with a not very definitive result, the first real detection  was achieved by \citet{BrainerdBlandfordS1996}. Mass measurements for galaxies in the Sloan Digital Sky Survey (SDSS) were achieved by \citet{McKayetal2002} and \citet {Pradaetal2003}. Dark matter was detected by \citet{KaiserSquires1993} and lensing produced by clusters of galaxies was detected by \citet{TysonValdesW1990}, and \citet{Fahlmanetal1994}. There have been a number of searches for spiral lens galaxies by \citet{Feronetal2009}, \citet{Sygnetetal2010} and \citet{Treuetal2011}.

The Hubble Space Telescope $(HST)$ has been used to take data on weak lensing galaxies for a few tens of ETGs with $0.1 \leq z \leq 0.8$  see \citep{Gavazzietal2007}, \citep{AugerTreuGetal2010}, \citep{Lagattutaetal2010}.

There is also what is called Strong Gravitational Lensing. There is a multitude of papers that have either tried to detect it or have detected it and in so doing, one of their byproducts has been the measurement of masses of clusters of galaxies. Papers such as those by \citet{BertinSagliaS1992}, \citet{SagliaBertinS1992}, \citet{LoewensteinWhite1999}, \citet{Keeton2001}, and \citet{Padmanabhanetal2004} to mention a few.

There have been a few strong lensing surveys such as SLACS (Sloan Lens ACS) \citet{Bolton2004}, \citet{Bolton2005}, \citet{Bolton2006}, \citet{Bolton2008},\citet{Schneider2006a}, \citet{Koopmansetal2006}, \citet{Treuetal2006}, \citet{Treuetal2009}, \citet{Gavazzietal2007},\citet{Treu2010}, \citet{AugerTreuetal2011} and \citet{Newtonetal2011} among others.

In order to apply the first two methods (dynamics, luminosity) it is necessary to know the distance at which the galaxy in question is located. The further away this galaxy is, the larger the uncertainties in the distance determination, and therefore the larger the uncertainties in its mass.

\end{enumerate}

In the case of LTGs, they contain different structures (bulge, disk, spiral arms) with different levels of importance that make it difficult to calculate the dynamical or the virial mass properly because the tracers of these structures are different. The best way to obtain the dynamical mass of LTGs is using the rotation curves but the present-day technology is not good enough to obtain these rotation curves for relatively distant LTGs. Besides the amount of time to obtain the rotation curves of LTGs is relatively high. At the present time, we can find in the literature photometric and spectroscopic information of thousands of galaxies from different surveys, but this information is not sufficient to obtain the rotation curves due to the previously mentioned problems. However, the SDSS contains more than one hundred thousand LTGs with photometric and spectroscopic information that can be used to obtain their virial masses through the velocity dispersion of gas and/or stars. This work is devoted to using this valuable information from the SDSS and obtaining virial mass estimations of LTGs with the lowest possible uncertainties. To achieve this goal we will use a subsample of SDSS LTGs with measured rotation curves from the literature to obtain their dynamical mass and compare it with the virial mass estimation from the velocity dispersion. This comparison will permit us to calibrate the virial masses, taking into account the importance of the different structures that constitute the LTGs.

This work is organized as follows: In section \ref{sec:sample} we present our sample of galaxies and the method by which it was selected. Section \ref{sec:masses} presents the formal definition of dynamic and virial masses. In section \ref{sec:virialmass} we calculate the virial masses of the galaxies in our sample. In section \ref{sec:dynamicalmass} we calculate the dynamical masses of a subsample of galaxies which have rotation curves determined in the literature. Section \ref{sec:calibration} presents the comparison of the virial and dynamical masses. In section \ref{sec:discussion} we discuss our findings and finally in section \ref{sec:conclusions} we present our conclusions.

\section{The sample of LTGs}
\label{sec:sample}

We extract a sample of LTGs -exponential brightness profile- from the SDSS DR16 \citep{yor00,bla03} with photometric and spectroscopic information. The basic required parameters to achieve the goals of this work are size and velocity dispersion from stars and gas. The main selection criterion used to quantify the brightness profile was the SDSS $fracDeV$ parameter. The $fracDeV$ parameter is equivalent to the \citet{Sersic1968} 
index $n$, $n$ = 1 corresponds to $fracDeV$ = 0 while $n$ = 4 corresponds to $fracDeV$ = 1. Galaxies with $fracDeV<0.5$ are relatively well adjusted by a exponential profile. Since the photometric errors are lower in $g$ and $r$ filters and the spectroscopic errors are lower for velocity dispersion greater than 60 $km/s$, we take into account this information to obtain the sample of LTGs. The resultant sample with $fracDeV_g<0.5$, $fracDeV_r<0.5$ and stellar $\sigma>$ 60 $km/s$ -according to SDSS nomenclature- contains 126 815 galaxies. These galaxies are distributed in a redshift interval $0.00 < z < 0.35$ and within a magnitude range $\Delta m\sim 10$ $mag$.

\subsection{Correction of the photometric and spectroscopic data}

Once we have compiled the galaxy sample, we must correct both the photometric and the spectroscopic data. We here list the performed corrections \citep[see][]{nig15}:

\begin{itemize}

\item Seeing and Extinction corrections: we use the seeing-corrected parameters in the total magnitude and effective radius as well as the extinction corrections by employing the respective SDSS pipelines on the data \citep[see][and references therein]{yor00,bla03}.

\item K correction: We use the values obtained from the following formulae \citep{nig08}:

\begin{equation}  \label{eq:eqs1}
k_g (z) =  -5.261z^{1.197},
\end{equation}

\begin{equation}  \label{eq:eqs2}
k_r (z) =  -1.271z^{1.023},
\end{equation}

\item Cosmological dimming correction: We follow \citet{jor95a} where the effective surface brightness ($<\mu_e>$) is corrected by subtraction of 10 log(1+$z$), being $z$ the redshift relative to the CMB.

\item Evolution correction: We apply the evolution correction from \citet{nig08} utilising:

\begin{equation}  \label{eq:eqs3}
 ev_g (z) =  +1.15z,
\end{equation}

\begin{equation}  \label{eq:eqs4}
ev_r (z) =  +0.85z,
\end{equation}

\item Effective radius correction to the rest reference frame: We follow \citet{hyd09} to correct colour gradients where the mean effective radius is smaller at longer wavelengths by using:

\begin{equation}
    r_{e,g,rest} = \left[\frac{(1+z)\lambda_g-\lambda_r}{\lambda_g-\lambda_r}\right] \left(r_{e,g,obs} - r_{e,r,obs}\right) + r_{e,r,obs},
\end{equation}

with $\lambda_g$ = 4686\AA, and $\lambda_r$ = 6165\AA.

\item Aperture correction to the velocity dispersion: The aperture correction is significant because it avoids dependencies on distance and instruments used in calculating the spectral parameters. Following \citet{jor95b}, we determine the ratio between the velocity dispersion values ($\sigma_{SDSS}$) and the corrected velocity dispersion ($\sigma_e$) or velocity dispersion inside r$_e$ as: 

\begin{equation}
    log\left(\frac{\sigma_{SDSS}}{\sigma_e}\right)= -0.065 log\left(\frac{r_{ap}}{r_e}\right) - 0.013\left[log\left(\frac{r_{ap}}{r_e}\right)\right]^2,
\end{equation}

where $r_{ap}$ = 1.5 arcsec for the SDSS case \citep[][]{yor00,bla03}.

\end{itemize}

Finally, the errors in the photometric and spectroscopic variables were obtained from the errors given in the SDSS, which were in turn propagated by considering the mathematical expressions of each of the corrections listed above.

\section{The mass estimations of the LTGs}
\label{sec:masses}

We assume Newtonian dynamics and that the velocity dispersion is due to the gravitational potential well. The LTGs mass estimates are made using data for stars and gas obtained from the SDSS DR16. These estimates will be calibrated later by means of a sample of galaxies for which we have measured rotation curves.

We have defined the virial masses as those masses that are obtained using the velocity dispersion of the gas or of the stars in the galaxies, while the dynamical masses would be those masses obtained by means of the rotation
curves. It is important to stress that this nomenclature is used as a means to differentiate the method by which the masses are obtained, however both mass-values are due to the gravitational potential of the galaxy in question and are derived using Newtonian dynamics.

\section{The virial mass from the stars of LTGs.}
\label{sec:virialmass}

 We have made seven estimates of the total virial mass (${\bf M_{Virial}}$) (see section 4.1) considering the following equation \citep{Poveda1958}:

\begin{equation}  \label{eq:eq2}
{\bf M_{Virial}} \sim  K\frac{ r_{e} \sigma_{e}^{2}}{G},
\end{equation}

where $r_{e}$ and, $\sigma_{e}$ represent the effective radius and the velocity dispersion of stars inside $r_{e}$, respectively, $G$ is the gravitational constant and $K$ is the proportion or scale factor. Historically, the scale factor has been considered constant, for example, for the de Vaucouleurs profile case \cite{cap06} found $K = 5.9$. Further studies have found the scale factor seems to depend on the Sersic index and/or the inclination angle of the galactic plane \citep{cap06,moc12}. In this work we will consider both constant and variable scale factor (see section 4.1). This scale factor will be calibrated with the analysis that we will perform in the following sections.

Equation \ref{eq:eq2} considers an idealized situation and does not take into account possible effects on the mass estimations of LTGs due to environment, shape and velocity dispersion anisotropies. 

\subsection{Calculation of virial masses considering different scale factors and different estimates of the velocity dispersion}

\label{sec:sec3.3}

 As previously mentioned the data for the virial mass estimates for the LTGs have been taken from the SDSS DR16. The data used are: the redshift, the exponential effective radius, the Petrosian radii R90 and R50, the major and minor semi-axes of the galactic disk, the stellar velocity dispersion, the gas velocity dispersion using the $H_{\alpha}$ line, and the gas velocity dispersion using the average of some of the Balmer lines ($H_{\alpha},H_{\gamma},H_{\beta},H_{\delta},H_{\epsilon},H_{\zeta},H_{\eta}$). In what follows we present these estimates.

\begin{enumerate}

    \item $\mathbf{M_{KS}}$. This mass was calculated using a constant scale factor ($K$=5.9) and the velocity dispersion of the stars.

    \item $\mathbf{M_{KA}}$. This mass was calculated using a constant scale factor ($K$=5.9) and the velocity dispersion of $H_{\alpha}$.

    \item $\mathbf{M_{KB}}$. This mass was calculated using a constant scale factor ($K$=5.9) and the average of the velocity dispersion of the Balmer lines.

    \item $\mathbf{M_{nS}}$. This mass was calculated using a scale factor as a function of the Sersic index $n$ ($K=8.87-0.831n+0.0241n^2$) \citep{cap06} and the velocity dispersion of the stars.

For the remaining three estimates we considered $K$ as a function of the galactic inclination angle ($i$) \citep{moc12} as expressed in the following equations:

\begin{equation}
    K=\left(\frac{2.3548}{sin(i)}\right)^2,
\end{equation}

\begin{equation}
    sin(i)=\sqrt{\frac{1-(\frac{b}{a})^2}{1-(0.19)^2}},
\end{equation}

where $a$ and $b$ represent the major and minor semi-axes of the galactic disk.

    \item $\mathbf{M_{IS}}$. This mass was calculated using a scale factor expressed as a function of the inclination angle (equation 9) and the velocity dispersion of the stars.

    \item $\mathbf{M_{IA}}$. This mass was calculated using a scale factor expressed as a function of the  inclination angle (equation 9) and the velocity dispersion of $H_{\alpha}$.

    \item $\mathbf{M_{IB}}$. This mass was calculated using a scale factor expressed as a function of the inclination angle (equation 9) and the average of the velocity dispersion of the Balmer lines.

\end{enumerate}

In table A.1 (appendix A) we show an extract (145 LTGs) of the virial masses catalogue, of the 126 815 LTGs, obtained following the previous procedures. The entire catalogue could be found, in electronic form, in the following link:

\section{Dynamical masses of a subsample of LTGs using rotation curves}
\label{sec:dynamicalmass}

 \cite{cat05} presented a long slit spectroscopic study in $H_{\alpha}$ of 403 non-interacting spiral galaxies using the Palomar Hale 5 m telescope. One of the main objectives of this study was to obtain their rotation curves. 
 In this work we selected all the galaxies from \citet{cat05} which have photometric and spectroscopic information in the SDSS DR16 sample, resulting in a subsample of 145 galaxies.  The spectroscopic information includes the stellar velocity dispersion and/or the gas velocity dispersion.
 
In what follows we present the details of the calculation of the dynamical mass for the 145 \cite{cat05} galaxies using their rotation curves. We make available the results in Table A2. 

In order to determine the dynamical masses of the LTGs, a mass component fitting was performed. The mass components were represented by a Miyamoto-Nagai Potential  \citep{miy75}:
 
 \begin{equation}
    {\phi}=\frac{GM}{\sqrt{R^{2}+{[a+{(b^2+z^2)}^{1/2}]}^2}},
\end{equation}

where $G$ is the universal gravitational constant and $M$ the total mass generating the gravitational field; $a$ and $b$ are two constants representing the scale length and scale height respectively, whereas $R$ and $z$ are the spatial variables.

A disk was set as the main component. Depending on the quality of the kinematic data, a second spherical inner component was included. Additionally, for some objects, an external spherical component was required to account for the flat region of the rotation curves at large radii. The obtained masses were corrected by inclination following the SDSS DR16 data. Respect to the fitting process, the main difficulties came from the deviation from circular motion which is present in almost all LTGs. Other aspects such as the spatial sampling, the radial extension of kinematic data and the rotational center determination were also considered as source of uncertainties. The obtained dynamical masses and their errors are listed in  Appendix A (Table A.2).

 If we only consider the information on galaxies with relatively small errors (less than 30\% in both virial and dynamical mass) and difference between virial and dynamical masses less than three times the dispersion (see Tables 1-6), the sample is reduced to a total of approximately 80 galaxies, with small variations in the total number depending on the subsample selection criteria. We used these subsamples in the following sections to perform the calibration of the virial masses.

\begin{table*}

\renewcommand{\footnoterule}{}  

\caption{ $BCES_{Bis}$ fit parameters for ${\bf M_{Virial}}$ vs. ${\bf M_{Dyn}}$ for LTG samples considering different scale factors and distinct estimates of the velocity dispersion (see Section 4.1). We apply approximately symmetric cuts in the values of absolute magnitude in the \textit{g}
filter M$_g$.}

\scalebox{0.60}{
\begin{tabular}{|l|l|c|c|c|c|c|}

\hline
\hline
                          &                             &                 &{\Huge  }         &{\Huge}        &  \\
                           &                             &                 &                        &               &  \\              
                         & & {\Huge Slope} &{\Huge Intercept}                          & {\Huge Correlation coefficient}  & {\Huge Dispersion} & {\Huge No. of galaxies}\\


                           &                             &                 &                                                          \\
\hline

        &                             &                 &                                                         \\

{\Huge \textit{M$_g$} $>$ -20.2}        &   & &    &               \\
                           &                             &                 &                                                          \\
\hline
                           &                             &                 &       &                                                   \\
\hline
\hline

\end{tabular}}
\end{table*}

\begin{table*}

\renewcommand{\footnoterule}{}  

\caption{ $BCES_{Bis}$ fit parameters for ${\bf M_{Virial}}$ vs. ${\bf M_{Dyn}}$ for LTG samples considering different scale factors and distinct estimates of the velocity dispersion (see Section 4.1). We apply approximately symmetric cuts in the values of Sersic index $n$.}

\scalebox{0.60}{
\begin{tabular}{|l|l|c|c|c|c|c|}

\hline
\hline
                          &                             &                 &{\Huge  }         &{\Huge}        &  \\
                           &                             &                 &                        &               &  \\              
                         & & {\Huge Slope} &{\Huge Intercept}                          & {\Huge Correlation coefficient}  & {\Huge Dispersion} & {\Huge No. of galaxies}\\
                           &                             &                 &                                                          \\
\hline

        &                             &                 &                                                         \\

{\Huge $n \leq 0.8$}        &   & &    &               \\
                           &                             &                 &                                                         \\
\hline
                           &                             &                 &       &                                                   \\
\hline
\hline

\end{tabular}}
\end{table*}

\begin{table*}

\renewcommand{\footnoterule}{}  

\caption{ $BCES_{Bis}$ fit parameters for ${\bf M_{Virial}}$ vs. ${\bf M_{Dyn}}$ for LTG samples considering different scale factors and distinct estimates of the velocity dispersion (see Section 4.1). We apply approximately symmetric cuts in the concentration parameter (R90/R50).}

\scalebox{0.60}{
\begin{tabular}{|l|l|c|c|c|c|c|}

\hline
\hline
                          &                             &                 &{\Huge  }         &{\Huge}        &  \\
                           &                             &                 &                        &               &  \\              
                         & & {\Huge Slope} &{\Huge Intercept}                          & {\Huge Correlation coefficient}  & {\Huge Dispersion} & {\Huge No. of galaxies}\\

                           &                             &                 &                                                          \\
\hline

        &                             &                 &                                                         \\

{\Huge (R90/R50) $\leq 2.4$}        &   & &    &               \\
                           &                             &                 &                                                         \\
\hline
                           &                             &                 &       &                                                   \\
\hline
\hline

\end{tabular}}
\end{table*}

\begin{table*}

\renewcommand{\footnoterule}{}  

\caption{ $BCES_{Bis}$ fit parameters for ${\bf M_{Virial}}$ vs. ${\bf M_{Dyn}}$ for LTG samples considering different scale factors and distinct estimates of the velocity dispersion (see Section 4.1). We apply approximately symmetric cuts in the values of color $g - r$.}

\scalebox{0.60}{
\begin{tabular}{|l|l|c|c|c|c|c|}

\hline
\hline
                          &                             &                 &{\Huge  }         &{\Huge}        &  \\
                           &                             &                 &                        &               &  \\              
                         & & {\Huge Slope} &{\Huge Intercept}                          & {\Huge Correlation coefficient}  & {\Huge Dispersion} & {\Huge No. of galaxies}\\

                           &                             &                 &                                                          \\
\hline

        &                             &                 &                                                         \\

{\Huge $g -r \leq 0.6$}        &   & &    &               \\
                           &                             &                 &                                                         \\
\hline
                           &                             &                 &       &                                                   \\
\hline
\hline

\end{tabular}}
\end{table*}


\begin{table*}

\renewcommand{\footnoterule}{}  

\caption{ $BCES_{Bis}$ fit parameters for ${\bf M_{Virial}}$ vs. ${\bf M_{Dyn}}$ for LTG samples considering different scale factors and distinct estimates of the velocity dispersion (see Section 4.1). We apply color cuts considering the lower limit of the red sequence ($\psi = −0.02667M_{r} + 0.11333 $) (Cooper et al. 2010).}

\scalebox{0.60}{
\begin{tabular}{|l|l|c|c|c|c|c|}

\hline
\hline
                          &                             &                 &{\Huge  }         &{\Huge}        &  \\
                           &                             &                 &                        &               &  \\              
                         & & {\Huge Slope} &{\Huge Intercept}                          & {\Huge Correlation coefficient}  & {\Huge Dispersion} & {\Huge No. of galaxies}\\

                           &                             &                 &                                                          \\
\hline

        &                             &                 &                                                         \\

{\Huge $g -r \leq \psi $}        &   & &    &               \\
                           &                             &                 &                                                         \\
\hline
                           &                             &                 &       &                                                   \\
\hline
\hline

\end{tabular}}
\end{table*}

\begin{table*}

\renewcommand{\footnoterule}{}  

\caption{ $BCES_{Bis}$ fit parameters for ${\bf M_{Virial}}$ vs. ${\bf M_{Dyn}}$ for LTG samples considering different scale factors and distinct estimates of the velocity dispersion (see Section 4.1). We apply approximately symmetric cuts in the values of inclination angle $i$.}


\scalebox{0.60}{
\begin{tabular}{|l|l|c|c|c|c|c|}

\hline
\hline
                          &                             &                 &{\Huge  }         &{\Huge}        &  \\
                           &                             &                 &                        &               &  \\              
                         & & {\Huge Slope} &{\Huge Intercept}                          & {\Huge Correlation coefficient}  & {\Huge Dispersion} & {\Huge No. of galaxies}\\

                           &                             &                 &                                                          \\
\hline

        &                             &                 &                                                         \\

{\Huge $i \leq 66$}        &   & &    &               \\
                           &                             &                 &                                                         \\
\hline
                           &                             &                 &       &                                                   \\
\hline
\hline

\end{tabular}}
\end{table*}

\section{Subsamples definition for the comparison of virial and dynamical masses}

The estimates of the virial and/or dynamical masses may be dependent on different properties, including the following: absolute magnitude, Sersic index, concentration index, colour ($g-r$) and galactic inclination angle. So, in order to perform the comparison in an appropriate manner we must consider subsamples with different cuts in the variables mentioned above. The estimation of the variable is described as follows:

\begin{itemize}

    \item The absolute magnitude was calculated considering the corrected apparent magnitude and the redshift from the SDSS.

    \item	The Sersic index was calculated through a fit to the $n$ vs. R90/R50 from Table 1 in \cite{gra05} in filters $g$ and $r$ from the SDSS, and later we took the average of these indices. R90 and R50 are the radii that contain 90\% and 50\% of the Petrosian flux, respectively.

    \item	The concentration parameter was obtained by averaging the ratio of the Petrosian radii (R90/R50) from the SDSS in the filters $g$ and $r$.

    \item The colour $g - r$ was obtained from the $g$ and $r$ corrected magnitudes in the SDSS.

    \item	The average inclination angle was calculated from equation 3 in filters $g$ and $r$ from SDSS.

\end{itemize}

\section{Calibration of the dynamical masses of the SDSS LTGs sample}
\label{sec:calibration}

In order to make the comparison between the dynamical and virial masses, we divide the samples in an approximately symmetrical way with respect to the different variables that might affect the estimates of the virial mass (see section 6). In Figure 1 we show a mosaic of the frequency distribution of the mentioned variables with the objective that the reader can visualize where the cuts were performed, dividing the sample approximately equally. Due to the different ways of calculating the masses and the number of variables investigated there are potentially more than 40 graphs that could be presented, but here we only show the graphs of one  of the biggest samples ($\mathbf{M_{nS}}$ sample) to avoid giving redundant information.

\begin{figure*}
   \begin{center}

      \includegraphics[angle=0,width=18cm]{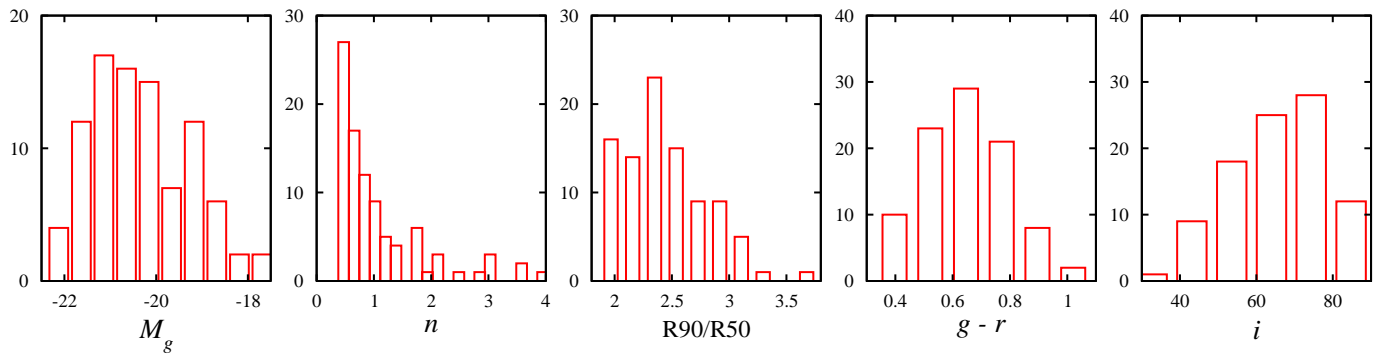}
      
         \caption{Frequency distribution of the variables that might affect the virial mass estimates for the $\mathbf{M_{nS}}$ sample. The approximately symmetrical cuts on the sample are made at -20.2, 0.8, 2.4, 0.6, 66$^{\circ}$ for magnitude, Sersic index, concentration, colour and inclination angle, respectively.}

         \label{FigVibStab4}
         \end{center}
   \end{figure*}

Once the cuts are made we perform a linear fit to the data. The coefficients of the fit could be affected significantly by both the choice of independent variables and the fitting method \citep{iso90}. The measurement errors in the variables may induce even bigger biases, as well as correlation in the errors or intrinsic dispersion in the relation. The Bivariate Correlated Errors and Intrinsic Scatter Bisector ($BCES_{Bis}$) method \citep{iso90,akr96,nig10} is a statistical model that takes into account the diﬀerent sources of bias mentioned above. Here, we use the $BCES_{Bis}$ method to calculate the coeﬃcients of the ${\bf M_{Virial}}$ vs. ${\bf M_{Dyn}}$ relationships. The results are as follows:

\begin{enumerate}

\item Absolute magnitude cut in the filter \textit{g} 
(\textit{M$_g$}).

The absolute magnitude cut was made at \textit{M$_g$} = -20.2. In Table 1 and Figure 2 we show the comparison between the virial and dynamical masses for the seven estimates of the virial mass with \textit{M$_g$} $>$ -20.2 and \textit{M$_g$} $\leq$ -20.2.

\item Sersic index cut.

The cut in the Sersic index \textit{n} was taken at \textit{n} = 0.8. In Table 2 and Figure 3 we show the results for the comparison of the virial and dynamical masses for the seven estimates of the virial mass with \textit{n} $>$ 0.8 and \textit{n} $\leq$ 0.8.

\item Concentration cut (R90/R50).

The cut in concentration was made at (R90/R50) = 2.4. In Table 3 and Figure 4 we show the results of comparing the virial and dynamical masses for the seven estimates of the virial mass with (R90/R50) $>$ 2.4 and (R90/R50) $\leq$ 2.4.

\item Colour \textit{g-r} cut.

The colour cut was taken in two different ways:

- Symmetrical cut in \textit{g-r}  = 0.6. In Table 4 and Figure 5 we show the results for the comparison of virial and dynamical masses for the seven estimates of the virial mass with \textit{g-r} $>$ 0.6 and \textit{g-r} $\leq$ 0.6.

- Knowing that the galaxies, including the spiral galaxies, are found in two well differentiated regions on a colour-magnitude diagram (the red sequence on the top part and the blue cloud on the bottom part; see Figure 6) we performed a cut in the colour \textit{g-r} considering the following lower limit to the red sequence ($\psi$) \citep{coo10}:

$\psi = −0.02667M_{r} + 0.113 33$

In Table 5 and in Figure 7 we show the results of the comparison of the virial and dynamical masses for the seven estimates of the virial mass with $g -r$ $> \psi$ and $g - r \leq \psi$.

\item Cut in inclination angle ($i$).

The cut in inclination angle was made at \textit{i} = 66$^{\circ}$. In Table 6 and Figure 8 we show the results of the comparison for the virial and dynamical masses for the seven estimates of the virial mass with \textit{i} $>$ 66$^{\circ}$ and \textit{i} $\leq$ 66$^{\circ}$.

\end{enumerate}

\section{Discussion}
\label{sec:discussion}

The comparisons of the virial and dynamical masses from the different samples of LTGs gave the following results:

\begin{itemize}

    \item When the magnitude is considered in the comparison of subsamples of LTGs we find that the correlation coefficients attain the lowest values and the fit dispersions attain the highest values with respect to the rest of the variables considered in the analysis (Sersic index $n$, concentration, colour, and angle of inclination).

There are neither substantial differences in the correlation fits nor in the dispersion when we compare samples of faint and bright galaxies (see Figure 2 and Table 1). The method used to make the estimate of the virial mass is of no consequence, that is to say, there are no significant differences whether the virial mass is obtained with a constant $K$ or if it is dependent on the Sersic $n$ index or the inclination angle. It also does not matter whether this virial mass was obtained using the velocity dispersions of the stars or of the gas.

    \item When we consider the Sersic index $n$, the correlation is relatively high in all fits (see Table 2). We can see that for $n \leq 0.8$ the slopes are lower than the slopes for $n > 0.8$. It is noticeable that for galaxies with $n \leq 0.8$ the correlation is lower when the virial mass is obtained using the stars compared to when this mass is obtained using the gas. For galaxies with $n > 0.8$ the opposite occurs. The method by which the virial mass is obtained turns out not to be significant for these results.

    \item When we consider the LTGs concentration parameter (R90/R50), we can see that for R90/R50 $<$ 2.4 the slopes are lower than the slopes for R90/R50 $>$ 2.4 (see Table 3). We also find a relatively high correlation for all fits. In this case, we observe that the more concentrated LTGs (R90/R50 $<$ 2.4) show a larger correlation, when we use the gas in estimating the virial mass, as opposed to when we use the stars. For the less concentrated galaxies (R90/R50 $>$ 2.4) the opposite occurs. The method by which the virial mass is calculated does not seem to be of much importance for these results.

    \item When the LTGs color is considered, we can see that for blue galaxies the slopes are lower than the slopes for red galaxies (see Tables 4 and 5). We also find a relatively high correlation in all cases. It can be noticed that for blue galaxies, the correlation is lower when the virial mass is obtained using the stars compared to when it is obtained by means of the gas. For red galaxies the opposite occurs. The method by which we obtained the virial mass does not seem to be of much consequence for the results above.

    \item When we take into account the inclination angle of the galaxy, we can see that for smaller inclination angles the slopes are greater than the slopes for larger inclination angles (see Table 6).  We also find a clear difference in correlation coefficient and dispersion for samples with smaller inclination angles as compared with samples with larger inclination angles. Again, the method utilized to estimate the virial mass appears not to be of much significance.

The correlation of the fits is larger (lower dispersion) for galaxies with inclination angles $i > 66^{o}$ both when the velocity dispersion of the stars or the velocity dispersion of the gas is used. How the virial mass estimate was made was not significant.

    \item In general, the fits show a lower correlation (larger dispersion) when the galaxies are separated by luminosity values and larger correlation (lower dispersion) when the galaxies are separated by inclination angle values (see Tables 1-6 and Figures 2-8). In the latter case, the method utilised to make the virial mass estimate is not important,  there are no significant differences whether the $K$ factor is taken as a constant or is dependent on the Sersic index $n$, or the inclination angle. Neither are there important differences if we use the velocity dispersion of the stars or the gas velocity dispersion for the virial mass estimates.

The estimates of virial masses given in this paper must be considered carefully and the results and suggestions given above should be taken into account in terms of the method and parameters used to obtain a better determination of the virial masses. In all cases, the best calibrations for the virial mass are those for which the inclination angle is relatively high ($i > 66^{o}$). On the other hand, if you have samples of spiral galaxies with one of the following properties; relatively small Sersic index, relatively small concentration or blue color, it is more appropriate to use the virial mass estimate obtained with the gas velocity dispersion ($H_{\alpha}$ or the average velocity dispersion of the Balmer lines). For the opposite cases it is better to use the estimate of the virial mass obtained with the velocity dispersion of the stars.

\end{itemize}

\section{Conclusions}
\label{sec:conclusions}
Due to the difficulty of calculating the dynamical masses of LTGs and 
with the objective of using the enormous amount of data from the SDSS, in the present paper we calculate different estimates of the virial mass for LTGs. The assumptions we use include Newtonian dynamics, and virial equilibrium. The data used for calculation are provided by the SDSS dynamical information of each galaxy as represented by its stars and gas. The virial masses were calibrated by comparing them with a subsample of galaxies whose dynamical masses were obtained using their rotation curves. Moreover, we characterized the possible effects that different variables might have on the values of the virial mass estimates. Among the main variables that could affect the estimates of the virial mass we find the following: the scale factor $K$ (see equation 8), the luminosity, the Sersic index $n$, the concentration index, the colour and the galaxy inclination angle.

The most important results that followed from the comparison of virial and dynamical masses, considering approximately symmetrical cuts in the variables mentioned above, are the following:

\begin{itemize}

\item In the case of cuts in luminosity no significant differences in the correlation coefficients of the fits for virial mass versus dynamical mass are found (see Table 1). We also find that the correlation coefficients are relatively low (relatively high dispersion), and that the method utilized to calculate the virial mass has no importance (see section 4.1). Neither are there important differences whether the stars or the gas are used in calculating the virial mass. From the above we conclude that galaxy luminosity is not an adequate variable to use to  improve the determination of the virial mass.

\item When we perform cuts in the Sersic index $n$, on the concentration index or on the colour $g-r$, the fits show a relatively high correlation in all cases (see Tables 2-4). We can see that for lower Sersic index, lower concentration and blue galaxies the slopes of the fits are lower than the slopes for greater Sersic index, greater concentration and red galaxies. It is interesting to note that if the Sersic index and/or the concentration are relatively low or if the color is blue, the correlation is higher if the velocity dispersion of the gas ($H_{\alpha}$ or the average velocity dispersion of the Balmer lines) are considered. For the opposite cases the correlation is higher if we consider the velocity dispersion of the stars. From this we may conclude that, as we might expect, for the smaller and bluer LTGs (dominated by the disk) the best indicator of the virial mass is the gas, while for larger and redder LTGs (dominated by the bulge), the best virial mass indicator is the stars. The method used to calculate the virial mass is of no consequence, in other words, there are no significant differences whether the virial mass is obtained with a constant $K$ or if it is dependent on the Sersic $n$ index or the inclination angle.

\item In the case of cuts in the galactic inclination angle we find a highest correlation in the fits (lower dispersion) as compared to the variables discussed in the previous two items. We also can see that for smaller inclination angles the slopes are greater than the slopes for larger inclination angles. An evidently higher correlation appears for samples of galaxies with a larger inclination angle. In this case no significant differences due to the method used to obtain the virial mass is found. Neither are there significant differences whether the stars or the gas are used in the estimate of the virial mass. From this we may conclude that the better determined virial masses (using indistinctly the gas or the stars) are those for which the inclination angle is relatively high and that the method used for calculating this mass is of no importance, therefore, we may use either a constant $K$ factor or a variable one without affecting substantially the estimate of the virial mass. The calibrations we used in this case are those presented in Table 6.

Considering these results and the calibration equation in Tables 1-6, we may conclude that the best calibrations for the virial masses of LTGs are those for galaxies with a relatively high inclination angle. If we have galaxies with relatively low inclination angles or we have samples of galaxies with heterogeneous properties one must take into consideration the size or colour of these galaxies through the use of the following variables: Sersic index $n$, concentration index or colour. For relatively smaller and bluer LTGs one must use the gas, while for LTGs that are relatively larger and redder one must use the stars.

If the reader wishes to perform their own virial mass estimates; they may use equation 1 and consider the calibration equations given in Tables 1-6 taking into consideration the properties of their sample and the advice given above.

\end{itemize}

\section*{Acknowledgements}

We thank the Instituto de Astronom\'{\i}a y Meteorolog\'{\i}a (UdG, M\'exico) and Instituto de Astronom\'{\i}a (UNAM, M\'exico) for all the facilities provided for the realisation of this project. A. Nigoche-Netro acknowledges support from CONACyT and PRODEP (M\'exico). Eduardo de la Fuente thanks Colegio departamental de F\'isica, authorities of CUCEI, Coordinaci\'on General Acad\'emica y de Innovaci\'on (CGAI-UDG), and the academic and administrative staff of the Institute for Cosmic Ray Research (ICRR), University of Tokyo (UTokyo), for authorization, acceptance, and support during his Sabbatical stay at ICRR in 2021. A. Ruelas-Mayorga thanks the Direcci\'on General de Asuntos del Personal Acad\'emico, DGAPA at UNAM for financial support under project numbers PAPIIT IN103813 and PAPIIT IN102617. P. Lagos is supported in the form of work contract (DL 57/2016/CP1364/CT0010) funded by national funds through Funda\c{c}\~ao para a Ci\^encia e Tecnologia (FCT), Portugal. We are indebted to Barbara Catinella who kindly made available the spiral galaxy rotation curves. Last but not least, we thank and acknowledge the comments made by an anonymous referee, which improved greatly the presentation of this paper.

\textbf{Data availability.} The data underlying this article are available in the article and in its online supplementary material.


\bibliographystyle{mnras}
\bibliography{DARKMATTERDENSIDAD}


\begin{figure*}
   \begin{center}

      \includegraphics[angle=0,width=14cm]{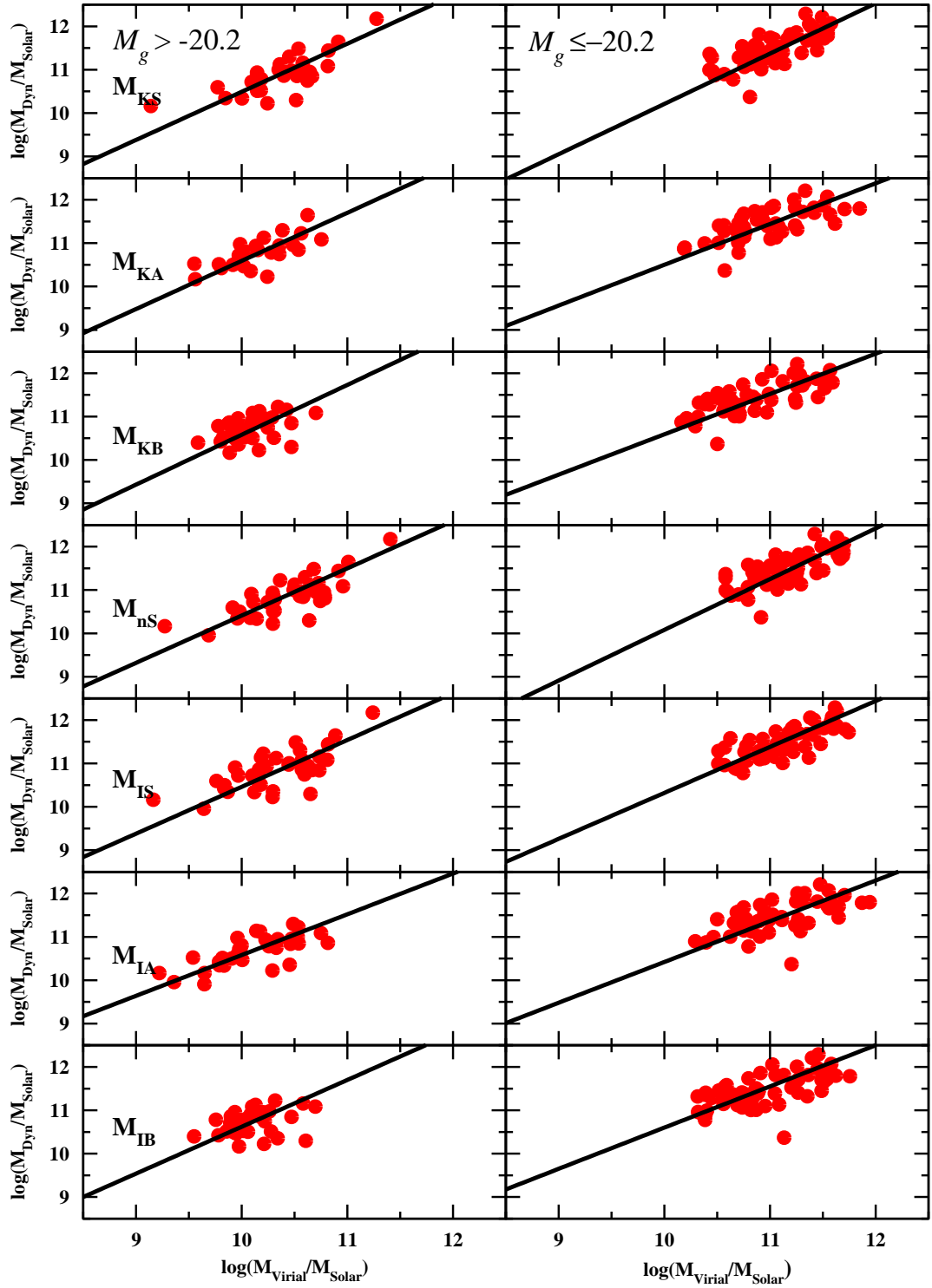}
      
         \caption{Distribution of the logarithmic difference between virial and dynamical mass for two absolute magnitude cuts ($M_{g} > -20.2$, $M_{g} \leq -20.2$) and different virial mass estimations. Black continuous lines are BCES$_{Bis}$ fits}

         \end{center}
   \end{figure*}

   
   \begin{figure*}
   \begin{center}

      \includegraphics[angle=0,width=14cm]{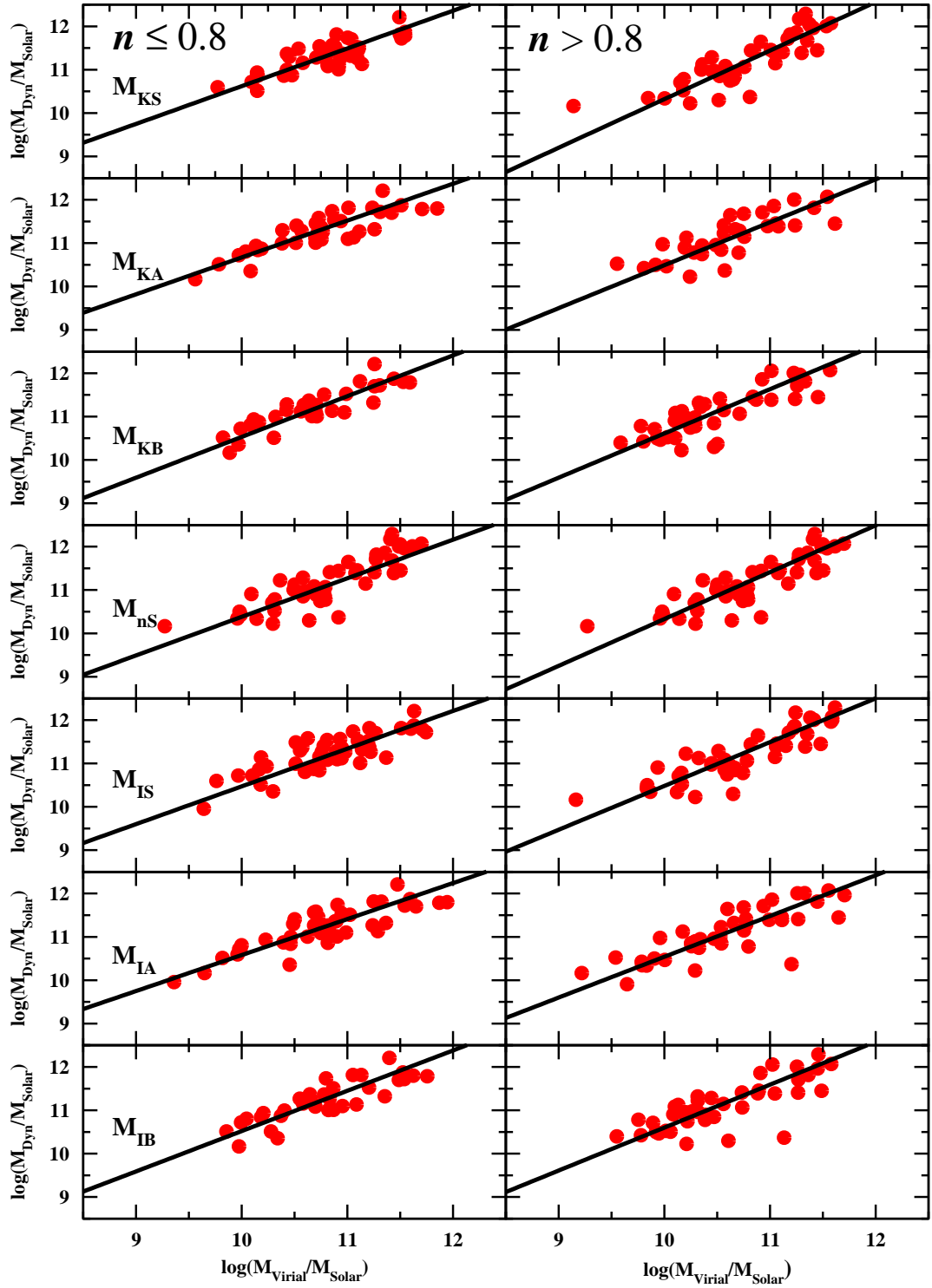}
      
         \caption{Distribution of the logarithmic difference between virial and dynamical mass for two Sersic index cuts ($n \leq 0.8$, $n > 0.8$) and different virial mass estimations. Black continuous lines are BCES$_{Bis}$ fits}

         \end{center}
   \end{figure*}
 

   \begin{figure*}
   \begin{center}

      \includegraphics[angle=0,width=14cm]{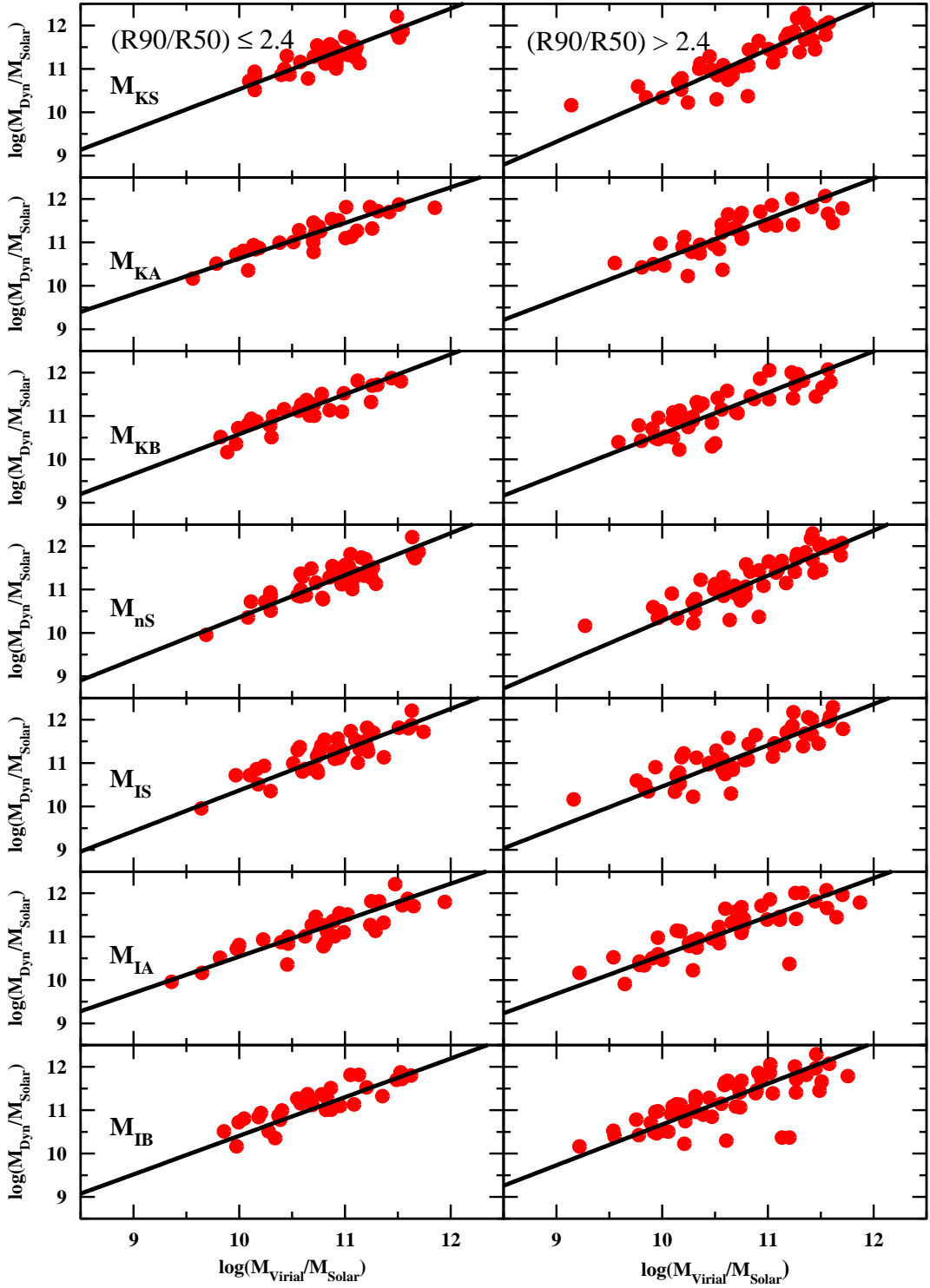}
 
         \caption{Distribution of the logarithmic difference between virial and dynamical mass for two concentration index cuts ($(R90/R50) \leq 2.4$, $(R90/R50) > 2.4$) and different virial mass estimations. Black continuous lines are BCES$_{Bis}$ fits}

         \end{center}
   \end{figure*}


   \begin{figure*}
   \begin{center}

      \includegraphics[angle=0,width=14cm]{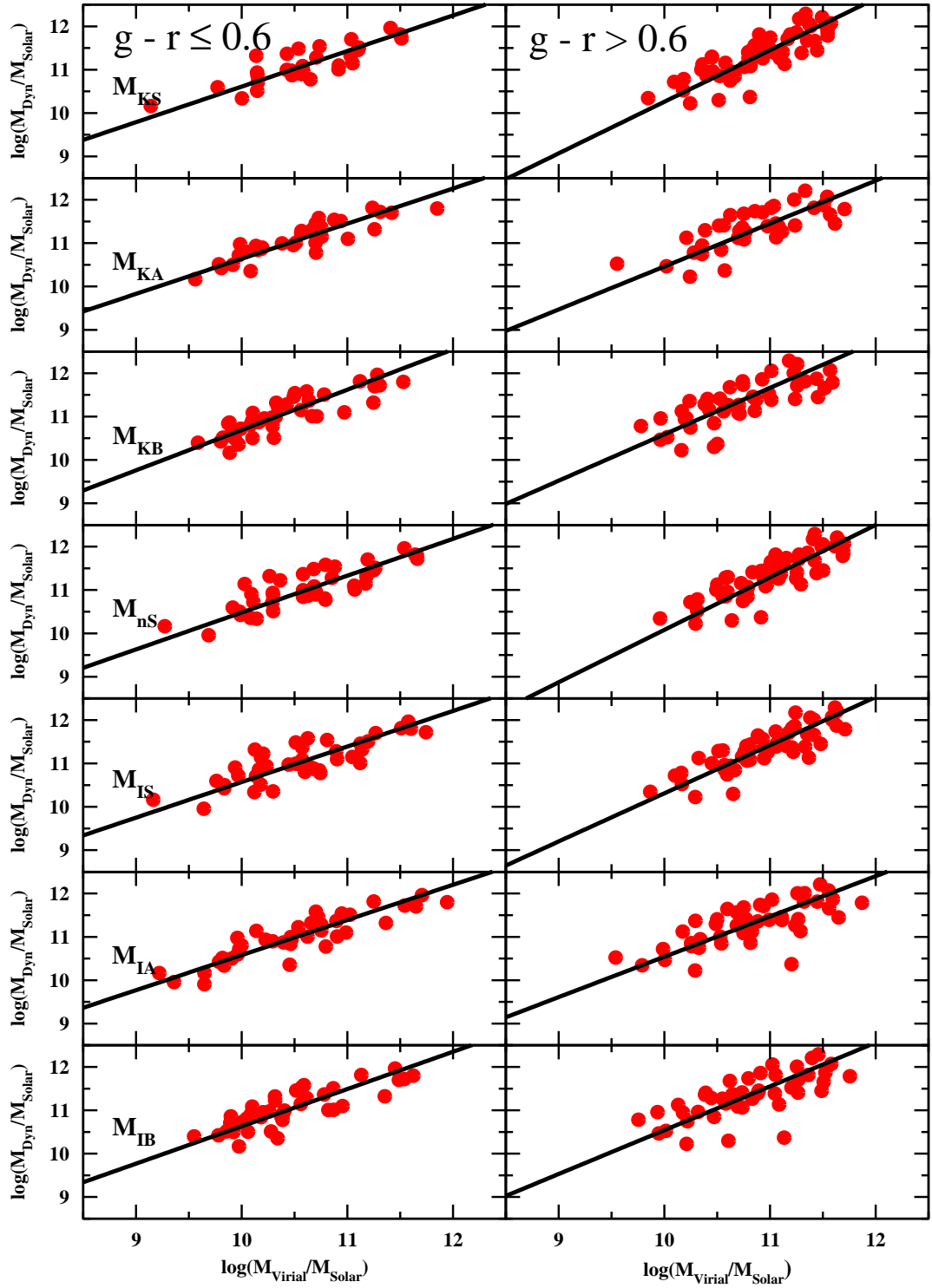}
      
         \caption{Distribution of the logarithmic difference between virial and dynamical mass for two colour cuts ($g - r \leq 0.6$, $g - r > 0.6$) and different virial mass estimations. Black continuous lines are BCES$_{Bis}$ fits}

         \end{center}
   \end{figure*}


\begin{figure*}
   \begin{center}

      \includegraphics[angle=0,width=10cm]{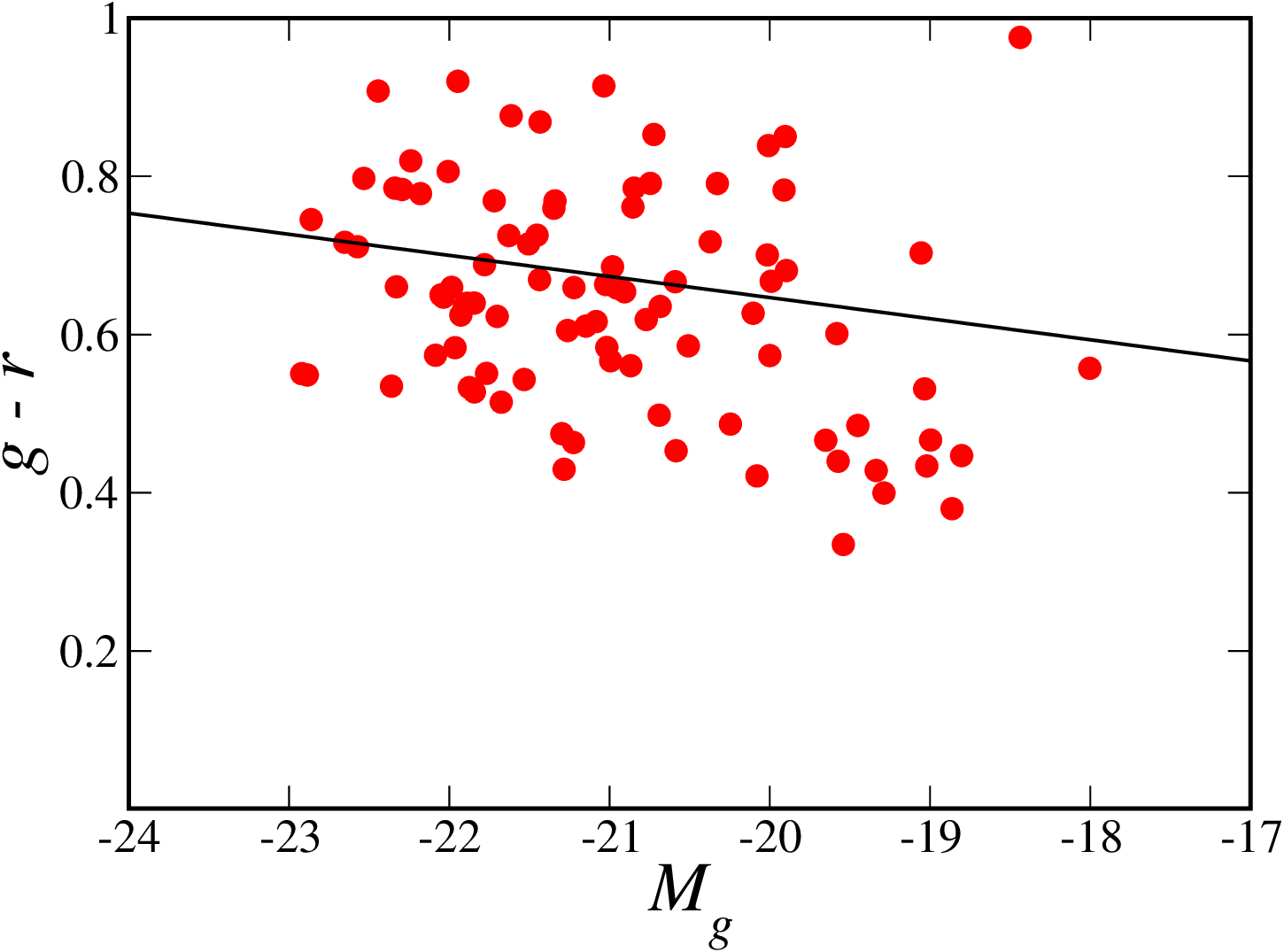}
      
         \caption{Colour-magnitude diagram of the $\mathbf{M_{nS}}$ sample. The black line represents the limit of the red sequence according to \protect\cite{coo10}.}

         \end{center}
   \end{figure*}


   \begin{figure*}
   \begin{center}

      \includegraphics[angle=0,width=14cm]{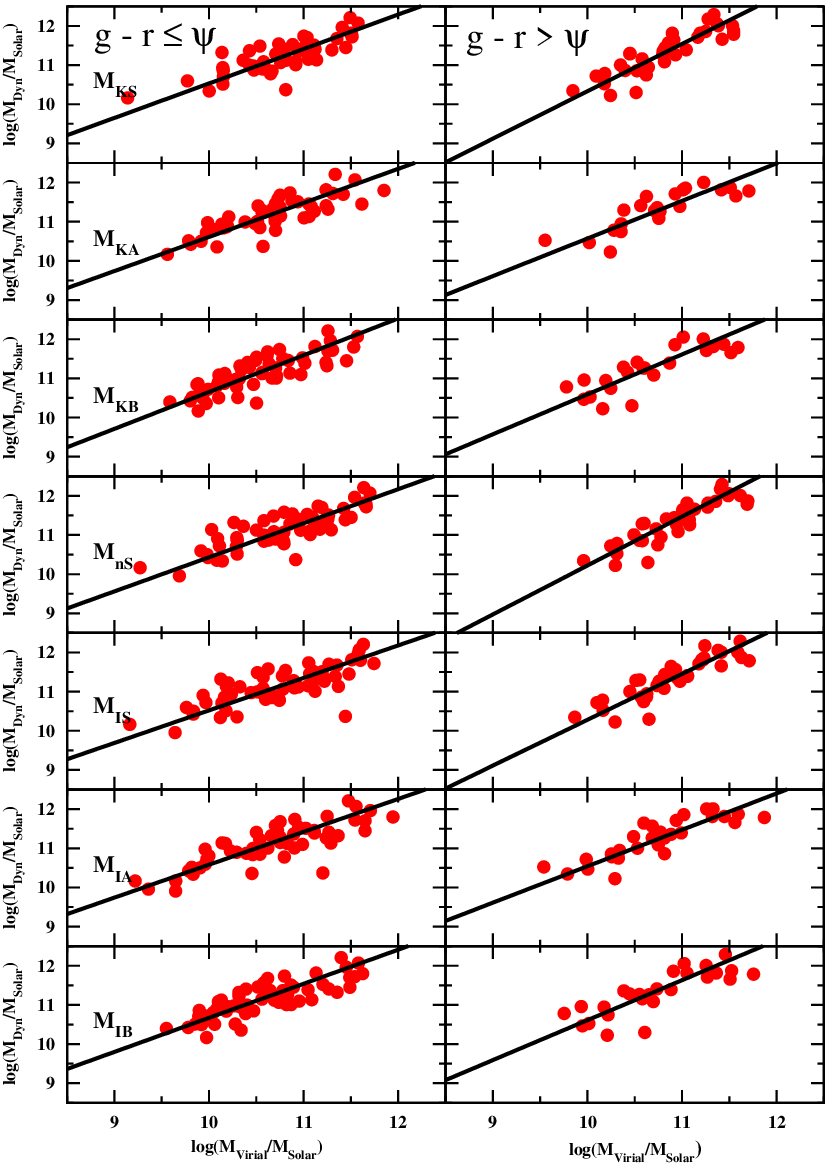}
      
         \caption{Distribution of the logarithmic difference between virial and dynamical mass for two colour cuts considering the Red Sequence limit ($\psi = −0.02667M_{r} + 0.113 33 $; \citealt{coo10}) 
         and different virial mass estimations. Black continuous lines are BCES$_{Bis}$ fits.}


         \end{center}
   \end{figure*}


   \begin{figure*}
   \begin{center}

      \includegraphics[angle=0,width=14cm]{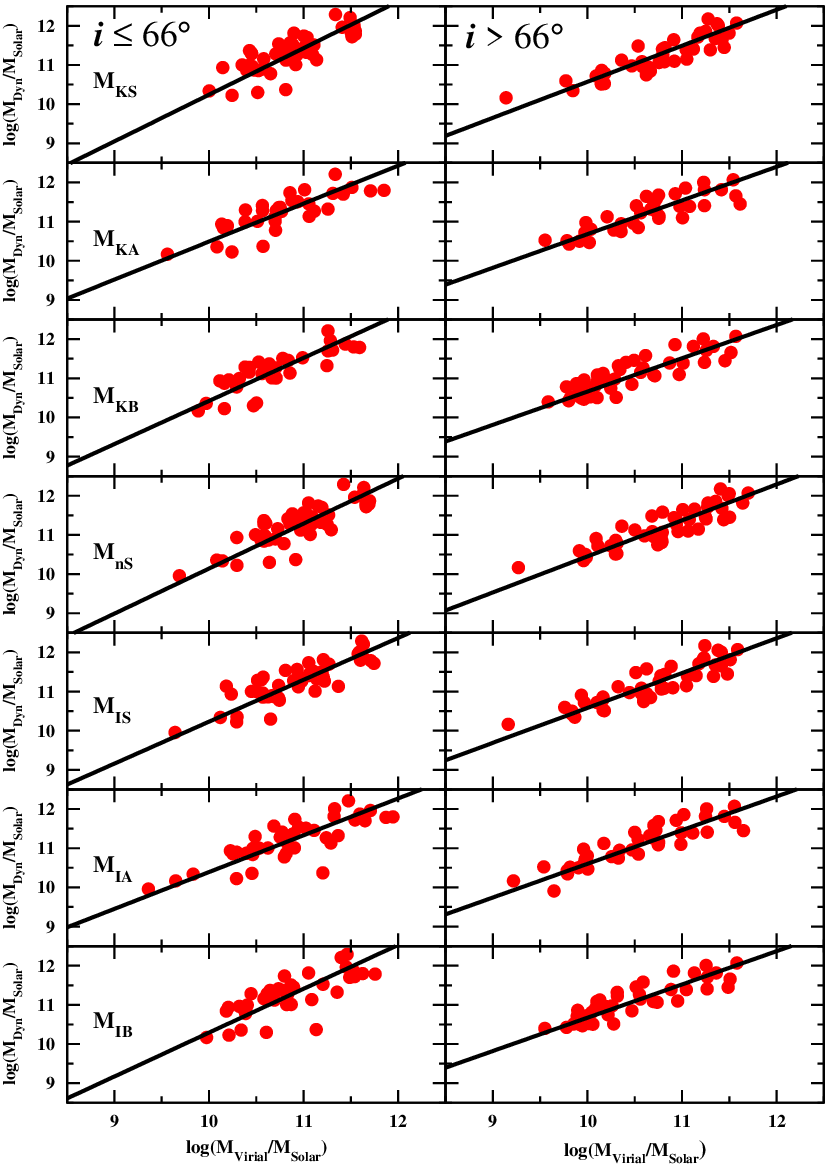}
      
         \caption{Distribution of the logarithmic difference between virial and dynamical mass for two inclination angle cuts ($i \leq 66^{o}$, $i > 66^{o}$) and different virial mass estimations. Black continuous lines are BCES$_{Bis}$ fits}

         \end{center}
   \end{figure*}

\clearpage

\section{Appendix A}

\setcounter{table}{0}
\renewcommand{\thetable}{A\arabic{table}}

In Table A.1 we show the virial masses, and some parameters involved in the analysis of this work, of a sub-sample of 145 LTGs from SDSS DR16 with information in \cite{cat05}. The virial masses were obtained following the procedure described in section 4. In column 1 we find the galaxy name in SDSS DR16. In column 2 we find the redshift from SDSS DR16. In column 3 we find the absolute magnitude in the $g$ filter, estimated using the apparent magnitude from the SDSS DR16. In column 4 we find the colour $g - r$ from the SDSS DR16. In columns 5 and 6 we find the ratio of the Petrosian radii R90 and R50 from the SDSS DR16 in the $g$ and $r$ filters respectively. In columns 7 and 8 we find the Sersic index $n$, in the $g$ and $r$ filters respectively, estimated using the procedure described in section 6. In column 9 and 10 we find the galactic inclination angles, in the $g$ and $r$ filters respectively, estimated using  equation 1 from section 4.1. Finally, in columns 11, 12, 13, 14, 15, 16, and 17 we find the logarithmic virial masses $\mathbf{M_{KS}}$, $\mathbf{M_{KA}}$, $\mathbf{M_{KB}}$, $\mathbf{M_{nS}}$, $\mathbf{M_{IS}}$, $\mathbf{M_{IA}}$, and $\mathbf{M_{IB}}$  respectively. The entire catalogue of virial masses of 126 699 LTGs from the SDSS (see section 2) with the same structure of Table A.1, can be found in electronic form in the following link: 

It is very important to remark that the previously mentioned masses must be corrected, depending on the necessities of the reader, using Tables 1-6 from section 7 and the advice given in section 9.

In Table A.2 we show the dynamical masses of a sub-sample of 145 LTGs from \cite{cat05} with photometric and/or spectroscopic information in the SDSS DR16. The dynamical masses were obtained using the rotation curves from Catinnella et al. 2005 as  described in section 5. In column 1 we find the galaxy name in the Arecibo General Catalog (AGC). In Column 2 we show other names of the galaxy (NGC, IC, CGCG or MCG). In column 3 we find the SDSS DR16 name. In columns 4 and 5 we find the right ascension and declination, respectively. Finally, in column 6 we find the dynamical mass in solar masses.

\clearpage


\begin{table}
   \centering
           \caption{Virial masses of a subsample of 145 LTGs from \protect\cite{cat05} with photometric and/or spectroscopic information in the SDSS DR16.}

   \begin{turn}{270}
    \resizebox{19cm}{!}{
\begin{tabular}{|c|c|c|c|c|c|c|c|c|c|c|c|c|c|c|c|c|c|c|c|c|}
    \hline \hline
     SDSS DR16 & z & $M_{g}$ & $g-r$ & $R50/R90_{g}$ & $R50/R90_{r}$  & $n_{g}$ & $n_{r}$ & $i_{g}$ & $i_{r}$ & ${\bf M_{KS}}$ & ${\bf M_{KA}}$ & ${\bf M_{KB}}$ & ${\bf M_{nS}}$ & ${\bf M_{IS}}$ & ${\bf M_{IA}}$ & ${\bf M_{IB}}$ \\
         \hline
         1237653653447180545 & 0.01872$\pm$0.00002 & -19.05$\pm$0.0065  & 0.08504$\pm$0.0085 & 2.595$\pm$0.2061 & 2.665$\pm$0.0945 & 0.9331$\pm$0.3671 & 1.068$\pm$0.1989 & 78.47$\pm$0.6191 & 84.43$\pm$1.043 &  10.18$\pm$0.0953 & 10.28$\pm$0.0297 & 9.779$\pm$0.0793 & 10.31$\pm$0.0953 & 10.16$\pm$0.0400 & 10.26$\pm$0.0125 & 9.756$\pm$0.0333\\
         1237652629638873182 & 0.03893$\pm$0.00002 & -20.15$\pm$0.0050  & 0.6190$\pm$0.0066 & 2.328$\pm$0.0655 &2.391$\pm$0.0434  & 0.5865$\pm$0.0589 & 0.6482$\pm$0.0462 &  80.71$\pm$1.326 & 83.48$\pm$1.142  & 10.54$\pm$0.0799 & 10.25$\pm$0.0453 & 10.16$\pm$0.0540 & 10.68$\pm$0.0799 & 10.51$\pm$0.0336 & 10.23$\pm$0.0191 & 10.13$\pm$0.0227 \\
         1237652947456360495 & 0.01312$\pm$0.00001 & -20.95$\pm$0.0032 & 0.7691$\pm$0.0045 &2.956$\pm$0.2660 & 3.013$\pm$0.3245 &  1.937$\pm$1.075 & 2.181$\pm$1.480 & 6824$\pm$0.2975 & 68.23$\pm$0.2784 & 11.19$\pm$0.0112 & 11.42$\pm$0.0643 & 11.33$\pm$0.0495 & 11.28$\pm$0.0112 & 11.22$\pm$0.0045 & 11.45$\pm$0.0254 & 11.36$\pm$0.0195\\
         1237652947993624616 & 0.02032$\pm$0.00001 & -19.86$\pm$0.0041 & 0.3786$\pm$0.0059 & - & - & 0.3278$\pm$11.87 & 0.3402$\pm$16.78 & 49.55$\pm$0.5539 & 49.83$\pm$0.5344 & 9.062$\pm$0.6737 & 9.975$\pm$0.0180 & 9.957$\pm$0.0236 & 9.219$\pm$0.6737 & 9.264$\pm$0.2182 & 10.18$\pm$0.0059 & 10.16$\pm$0.0077\\
         1237652900753244294 & 0.01925$\pm$0.00001 & -18.89$\pm$0.0042 & 0.3997$\pm$0.0059 & 2.592$\pm$0.1168 &2.586$\pm$0.1665 & 0.9280$\pm$0.2066 & 0.9170$\pm$0.2899 & 80.57$\pm$0.9919 & 80.70$\pm$1.093 &  9.856$\pm$0.2327 & 9.808$\pm$0.0097 & 9.802$\pm$0.0140 & 9.989$\pm$0.2327 & 9.834$\pm$0.0975 & 9.786$\pm$0.0042 & 9.780$\pm$0.0059\\
\hline
    \end{tabular}}
    \end{turn}
\end{table}

\clearpage
\begin{table}
   \centering
   \begin{turn}{270}
    \resizebox{19cm}{!}{
    \begin{tabular}{|c|c|c|c|c|c|c|c|c|c|c|c|c|c|c|c|c|c|c|c|c|}
    \hline \hline
     SDSS DR16 & z & $M_{g}$ & $g-r$ & $R50/R90_{g}$ & $R50/R90_{r}$  & $n_{g}$ & $n_{r}$ & $i_{g}$ & $i_{r}$ & ${\bf M_{KS}}$ & ${\bf M_{KA}}$ & ${\bf M_{KB}}$ & ${\bf M_{nS}}$ & ${\bf M_{IS}}$ & ${\bf M_{IA}}$ & ${\bf M_{IB}}$  \\
         \hline

         1237650760241774845 & 0.05816$\pm$0.00002 & -20.53$\pm$0.0070 & 0.4246$\pm$0.0096 & 1.866$\pm$0.0565 & 2.066$\pm$0.0645 & 0.3571$\pm$0.0126 & 0.4197$\pm$0.0273 & 57.98$\pm$1.2501 & 57.80$\pm$1.145 & 9.673$\pm$1.392 & 10.51$\pm$0.0764 & 10.71$\pm$0.0854 & 9.827$\pm$1.392 & 9.783$\pm$0.5006 & 10.62$\pm$0.0276 & 10.82$\pm$0.0308\\
         1237650760241905743 & 0.07072$\pm$0.00001 & -21.67$\pm$0.0040 & 0.6604$\pm$0.0052 & 2.384$\pm$0.0543 & 2.500$\pm$0.0445 & 0.6408$\pm$0.0567 & 0.7816$\pm$0.0626 & 67.82$\pm$0.6083 & 67.67$\pm$0.5337 & 11.30$\pm$0.0340 & 11.08$\pm$0.0203 & 11.01$\pm$0.0212 & 11.44$\pm$0.0340 & 11.33$\pm$0.0134 & 11.11$\pm$0.0081 & 11.05$\pm$0.0084\\
         1237650760241905753 & 0.00525$\pm$0.00001 & -15.32$\pm$0.0125 & 0.4579$\pm$0.0179 & 2.148$\pm$0.0569 & 2.165$\pm$0.6043 & 0.4594$\pm$0.0308 & 0.4687$\pm$0.3437 & 83.70$\pm$1.944 & 81.25$\pm$1.779 & 9.550$\pm$0.4342 & - & - & 9.702$\pm$0.4342 & 9.524$\pm$0.1828 & 8.988$\pm$0.0320 & -\\
         1237650760241971341 & 0.02725$\pm$0.00001 & -20.12$\pm$0.0039 & 0.9143$\pm$0.0051 & 2.893$\pm$0.1370 & 2.940$\pm$0.1295 & 1.699$\pm$0.4836 & 1.874$\pm$0.5064 & 74.28$\pm$0.6195 & 75.71$\pm$0.4783 & 10.82$\pm$0.0287 & 10.17$\pm$0.0570 & 10.11$\pm$0.0591 & 10.92$\pm$0.0287 & 10.82$\pm$0.0118 & 10.17$\pm$0.0234 & 10.11$\pm$0.0243\\
         1237650369407090853 & 0.02666$\pm$0.00001 & -20.47$\pm$0.0036 & 0.6164$\pm$0.0049 & 2.677$\pm$0.2517 & 2.739$\pm$0.2183 & 1.093$\pm$0.5445 & 1.240$\pm$0.5465 & 75.61$\pm$0.4543 & 74.20$\pm$0.4063 & 11.05$\pm$0.0265 & 10.77$\pm$0.0347 & 10.56$\pm$0.0392 & 11.17$\pm$0.0265 & 11.05$\pm$0.0109 & 10.75$\pm$0.0144 & 10.56$\pm$0.0161\\
\hline
    \end{tabular}}
    \end{turn}
\end{table}

\clearpage
\begin{table}
   \centering
        \caption{Dynamical masses of a subsample of 145 LTGs from \protect\cite{cat05} with photometric and/or spectroscopic information in the SDSS DR16.}
    \resizebox{16cm}{!}{
    \begin{tabular}{|c|c|c|c|c|c|}
    \hline \hline
      AGC Name & Other Name & SDSS DR16 Name & $\alpha$ & $\delta$ & $\mathbf{M_{Dyn}}$  \\
         &  &	& J2000	& J2000 & $10^{11} {\bf M_{solar}}$  \\

        \hline
       
        144 & 456-044 &	1237653653447180545	& 00 15 26.8 & ~16 14 05 & 0.61$\pm$0.17\\
        400814 & - & 1237652629638873182 & 00 31 56.7 & -10 15 08 & 3.05$\pm$0.51\\
        400343 & N217 & 1237652947456360495 & 00 41 33.7 & -10 01 18 & 6.51$\pm$ 1.06\\
        400373 & M-203006 & 1237652947993624616 & 00 45 08.8 & -09 37 53 & 0.46$\pm$0.15\\
        400846 & M-203014 & 1237652900753244294 & 00 47 25.2 & -09 51 10 & 0.27$\pm$0.07\\
        400390 & M-203015 & 1237652900753244175 & 00 47 46.2 & -09 50 05 & 2.46$\pm$0.48\\
\hline
    \end{tabular}}
\end{table}

\clearpage

\begin{table}
   \centering
    \resizebox{17cm}{!}{
    \begin{tabular}{|c|c|c|c|c|c|}
    \hline \hline
      AGC Name & Other Name & SDSS DR16 Name & $\alpha$ & $\delta$ & ${\bf M_{Dyn}}$  \\
         &  &	& J2000	& J2000 & $10^{11} {\bf M_{solar}}$  \\

        \hline
       
        180869 & - & 1237661086957568117 & 08 27 13.3 & ~25 58 01 & 3.80$\pm$0.91\\
        4501 & N2620 & 1237664092898852896 & 08 37 28.2 & ~24 56 47 & 11.74$\pm$1.93\\
        4691 & N2713 & 1237654601019686930 & 08 57 20.6 & ~02 55 14 & 19.48$\pm$3.57\\
        4698 & 150-053 & 1237664093438214204 & 08 58 32.5 & ~28 15 59 & 11.81$\pm$2.00\\
        4823 & N2777 & 1237660412649865229 & 09 10 41.7 & ~07 12 23 & 0.14$\pm$0.05\\
        190864 & - & 1237671260664168763 & 09 15 28.2 & ~11 50 05 & 0.15$\pm$0.05\\
\hline
    \end{tabular}}
\end{table}

\clearpage

\begin{table}
   \centering
    \resizebox{17cm}{!}{
    \begin{tabular}{|c|c|c|c|c|c|}
    \hline \hline
      AGC Name & Other Name & SDSS DR16 Name & $\alpha$ & $\delta$ & ${\bf M_{Dyn}}$  \\
         &  &	& J2000	& J2000 & $10^{11} {\bf M_{solar}}$  \\

        \hline

        7297 & N4226 & 1237661433771261955 & 12 16 26.2 & ~47 01 32 & 2.52$\pm$0.56\\
        7337 & 013-121 & 1237648702966988820 & 12 18 08.6 & -01 03 52 & 0.91$\pm$0.16\\
        221991 & - & 1237665441511964722 & 12 19 44.0 & ~29 34 49 & 1.85$\pm$0.48\\
        7478 &  N4357 & 1237661149765107752 & 12 23 58.7 & ~48 46 45 & 3.67$\pm$0.81\\
        221659 & - & 1237661971186188446 & 12 25 41.7 & ~07 10 01 & 0.28$\pm$0.09\\
        221928 & - & 1237667255627087966 & 12 26 51.7 & ~31 05 39 & 6.11$\pm$1.75\\
\hline
    \end{tabular}}
\end{table}

\clearpage

\begin{table}
   \centering
    \resizebox{17cm}{!}{
    \begin{tabular}{cccccc}
    \hline \hline
     AGC Name & Other Name & SDSS DR16 Name & $\alpha$ & $\delta$ & ${\bf M_{Dyn}}$  \\
         &  &	& J2000	& J2000 & $10^{11} {\bf M_{solar}}$  \\

        \hline

        9645 & N5806 & 1237651736313528355 & 15	00 00.3 & ~01 53 28 & 1.01$\pm$0.22\\
        9681 & M+731023 & 1237661211510833280 & 15 03 39.7 & ~42 07 33 & 0.32$\pm$0.07\\
        9888 & I1125 & 1237655692481986636 & 15 33 05.6 & -01 37 42 & 0.22$\pm$0.05\\
        10306 & 023-031 & 1237655550742298700 & 16 16 43.5 & ~00 14 46 & 10.15$\pm$1.52\\
        10692 & 139-034 & 1237662302453039177 & 17 05 28.0 & ~23 09 11 & 4.76$\pm$0.73\\
        10706 & 197-035 & 1237654954281468455 & 17 06 28.3 & ~38 21 47 & 1.83$\pm$0.30\\
        600138 & M-153012 & 1237652598487646631	& 20 49 52.0 & -07 01 17 & 4.41$\pm$0.75\\
        11649 & 374-028 & 1237656567038935144 & 20 55 27.6 & -01 13 31 & 1.44$\pm$0.34\\
        310030 & 426-015 & 1237652944768008329 & 21 12 01.2 & ~11 29 45 & 0.20$\pm$0.05\\
        610024 & 375-011 & 1237663457240089137 & 21 13 12.8 & ~00 21 58 & 0.86$\pm$0.15\\
        11712 & N7047 & 1237663456703545733 & 21 16 27.6 & ~00 49 36 & 2.28$\pm$0.49\\
        610495 & - & 1237666210347614532 & 21 56 49.5 & -07 45 33 & 1.45$\pm$0.45\\
        610499 & - & 1237666210347745660 & 21 58 05.8 & -07	50 43 & 0.73$\pm$0.16\\
        12174 & N7364 & 1237663543150837809 & 22 44	24.2 & ~00 09 44 & 2.58$\pm$0.39\\
        12363 & 379-034 & 1237663784191524949 & 23 06 31.5 & ~00 10 18 & 0.99$\pm$0.20\\
        35971 & 604G7 & 1237672815977496655 & 23 15 17.2 & -21 21 57 & 3.18$\pm$0.85\\
        12834 & N7782 & 1237669518514913516 & 23 53 54.0 & ~07 58 14 & 9.73$\pm$1.59\\
    \hline
    \end{tabular}}
\end{table}









\bsp	
\label{lastpage}
\end{document}